\documentclass[conference]{IEEEtran}
\IEEEoverridecommandlockouts
\usepackage{cite}
\usepackage{amsmath,amssymb,amsfonts}
\usepackage{algorithmic}
\usepackage{graphicx}
\usepackage{subfigure}
\usepackage{textcomp}
\usepackage{xcolor}
\usepackage{multirow}
\usepackage{multirow}
\usepackage{booktabs}

\usepackage{algorithm} 
\usepackage{algorithmic} 
\usepackage{hyperref}

\def\BibTeX{{\rm B\kern-.05em{\sc i\kern-.025em b}\kern-.08em
    T\kern-.1667em\lower.7ex\hbox{E}\kern-.125emX}}
\begin{document}

\title{DGTN: Dual-channel Graph Transition Network for Session-based Recommendation}

% \author{\IEEEauthorblockN{Yujia Zheng}
% \IEEEauthorblockA{\textit{University of Electronic Science and Technology of China} \\
% % \textit{University of Electronic Science and Technology of China}\\
% Chengdu, China \\
% yjzheng19@gmail.com}
% \and
% \IEEEauthorblockN{Siyi Liu}
% \IEEEauthorblockA{\textit{University of Electronic Science and Technology of China} \\
% % \textit{University of Electronic Science and Technology of China}\\
% Chengdu, China \\
% ssui.liu1022@gmail.com}
% \and
% \IEEEauthorblockN{Zekun Li}
% \IEEEauthorblockA{\textit{Institute of Information Engineering, Chinese Academy of Sciences} \\
% \textit{School of Cyber Security, University of Chinese Academy of Sciences}\\
% Beijing, China \\
% lizekunlee@gmail.com}
% \and
% \IEEEauthorblockN{Shu Wu}
% \IEEEauthorblockA{\textit{Institute of Automation, Chinese Academy of Sciences} \\
% \textit{University of Chinese Academy of Sciences}\\
% Beijing, China \\
% shu.wu@nlpr.ia.ac.cn}

% }

\author{
    \IEEEauthorblockN{Yujia Zheng\IEEEauthorrefmark{1}, Siyi Liu\IEEEauthorrefmark{1}, Zekun Li\IEEEauthorrefmark{2}\IEEEauthorrefmark{3}, Shu Wu\IEEEauthorrefmark{4}\IEEEauthorrefmark{5}}
    \IEEEauthorblockA{\IEEEauthorrefmark{1}University of Electronic Science and Technology of China
}
    \IEEEauthorblockA{\IEEEauthorrefmark{2}School of Cyber Security, University of Chinese Academy of Sciences
    \\\IEEEauthorrefmark{3}Institute of Information Engineering, Chinese Academy of Sciences
}
    \IEEEauthorblockA{\IEEEauthorrefmark{4}School of Artificial Intelligence, University of Chinese Academy of Sciences
    \\\IEEEauthorrefmark{5}Institute of Automation and Artificial Intelligence Research, Chinese Academy of Sciences
    \\\{yjzheng19, ssui.liu1022, lizekunlee\}@gmail.com, shu.wu@nlpr.ia.ac.cn}

}

\maketitle

\begin{abstract}
The task of session-based recommendation is to predict user actions based on anonymous sessions. Recent research mainly models the target session as a sequence or a graph to capture item transitions within it, ignoring complex transitions between items in different sessions that have been generated by other users. These item transitions include potential collaborative information and reflect similar behavior patterns, which we assume may help with the recommendation for the target session. In this paper, we propose a novel method, namely Dual-channel Graph Transition Network (DGTN), to model item transitions within not only the target session but also the neighbor sessions. Specifically, we integrate the target session and its neighbor (similar) sessions into a single graph. Then the transition signals are explicitly injected into the embedding by channel-aware propagation. Experiments on real-world datasets demonstrate that DGTN outperforms other state-of-the-art methods. Further analysis verifies the rationality of dual-channel item transition modeling, suggesting a potential future direction for session-based recommendation.
\end{abstract}

\begin{IEEEkeywords}
Session-based Recommendation, Graph Neural Network
\end{IEEEkeywords}

\section{Introduction}
Session-based Recommendation System (SRS) has attracted much attention for its highly practical value, especially in some real-world scenarios that concentrated with multitudes of anonymous interactive data. Different from most of the other recommendation tasks that need explicit user demographic profiles, SRS only relies on anonymous user action logs (e.g., clicks) in an ongoing session to predict the user's next action. 

Under these circumstances, several methods are proposed to tackle the SRS task. Markov Chains (MC) \cite{rendle2010factorizing} is a representation of traditional methods. It predicts the user’s next action based on the previous one thus introduces sequentiality into SRS. 
Recently, neural network-based methods have become popular due to their strong abilities to model sequential data, such as the methods based on Recurrent Neural Networks (RNN)~\cite{hidasi2015session, tan2016improved, li2017neural}.
Unfortunately, they can only model the unidirectional transitions between consecutive items but neglect those among other contextual items in the same session.  
To solve that, SR-GNN~\cite{wu2019srgnn} models the target session in the graph structure and utilizes Graph Neural Networks (GNN) to model complex transitions among item nodes on the graph.
% FGNN~\cite{qiu2019rethinking} proposes a weighted graph attention layer and readout function to emphasize the inherent order of item transition pattern.
Despite the success of these methods based on RNN or GNN, they only focus on the item transitions within the target session but ignore those in the neighbor sessions.
As a result, they are deficient in modeling the complex item transitions among different sessions, which contains potential fine-grained collaborative information for prediction.

In this paper, we propose a novel method, namely Dual-channel Graph Transition Network (DGTN), to model item transitions within not only the target session but also its neighbor ones. 
We first construct the target session and its neighbor sessions into a single graph.
%Different from aggregating neighbor nodes indiscriminately, channel-aware information propagation leverages intra- and inter-session channels to deal with the differences between two types of transitions. 
To consider the difference between the item transitions within the target sessions and the neighbor sessions, we leverage intra- and inter-session channels for embedding propagation, respectively.
Then we use fusion function to aggregate features from the two channels.

The main contributions of our work are summarized as follows:
\begin{itemize}
    \item {We propose a novel model DGTN, which explicitly exploits item transitions among different sessions by constructing the target and neighbor sessions as a single graph and propagating embeddings on it in a channel-aware manner. }  
    \item {We evaluate our model on two real-world datasets. Extensive experiments demonstrate the state-of-the-art performance of DGTN and its rationality of explicitly modeling dual-channel item transitions.}
\end{itemize}

The rest of this paper is structured as follows. In $\S$ \ref{r} we introduce related works about session-based recommendation and graph embedding. In $\S$ \ref{p} we describe the necessary preliminaries. In $\S$ \ref{m} we elaborate the proposed model. The experiments and analysis are presented in $\S$ \ref{e}. And the concluding takeaways are in $\S$ \ref{c}.

\section{Related Work}
\label{r}
\noindent In this section, we introduce some related works in Session-based Recommendation, Collaborative Filtering, and Graph Embedding.

\subsection{Session-based Recommendation}
% \noindent Traditional methods for session-based recommendation are mainly based on MF \cite{koren2009matrix}, which uses the factorization of the user-item rating matrix to get general representations of users' preferences. Another widely used approach is the neighborhood-based method, which produces recommendations based on similarities between the last item in the current session and items in other sessions. Later works began to consider the sequentiality in SRS by introducing MC to predict user's next action based on the last action \cite{zimdars01,shani2005mdp,rendle2010factorizing}. However, MC-based methods lose a balance between user's general preference and sequential behavior, for they seldom consider sequentiality between items that are not consecutively adjacent in the same session. To achieve that balance, Rendle $\textit{et al.}$ \cite{rendle2010factorizing} propose a hybrid method taking account of the combination of MF and MC, namely FPMC.

\noindent Traditional methods for session-based recommendation are mainly based on Markov Chains (MC) \cite{zimdars01,Mobasher02,shani2005mdp,rendle2010factorizing}, which introduces the sequentiality in SRS by predicting the user's next action based on the last action. Zimdars $\textit{et al.}$ \cite{zimdars01} apply probabilistic decision-tree models to study the way to extract the sequentiality. Mobasher $\textit{et al.}$ \cite{Mobasher02} choose the contiguous sequential patterns for SRS after studying the effect of different patterns. Shani $\textit{et al.}$ \cite{shani2005mdp} employ the Markov Decision Processes that consider the long-term effect and the expected value of each recommendation. However, MC-based methods lose a balance between user's general preference and sequential behavior, for they seldom consider sequentiality between items that are not consecutively adjacent in the same session. To achieve that balance, Rendle $\textit{et al.}$ \cite{rendle2010factorizing} propose a hybrid method taking account of the combination of Matrix Factorization and MC, namely FPMC.

Like most other fields \cite{li2019fi, zhang2020every, hu2019graphair, zhu2020deep, chen2020jit2r}, deep learning methods \cite{yu2020tagnn, mi2020ader, liu2020long, wang2019modeling, wang2019survey, wang2020graph, wang2020intention, wang2020intention2basket, wu2019personalizing, chen2018sequential, ye2020time} frequently appear in recent SRS models and obtain new state-of-the-art performance in terms of accuracy, especially RNN-based methods \cite{hidasi2015session,tan2016improved,li2017neural,liu2018stamp}. Hidasi $\textit{et al.}$ \cite{hidasi2015session} employ RNN with the Gated Recurrent Unit (GRU) into SRS and outperform traditional methods. Tan $\textit{et al.}$ \cite{tan2016improved} further improve it by introducing data augmentation, distillation integrating privileged information, and a pre-training approach to account for temporal shifts in the data distribution. Later, attention mechanism is applied by an encoder-decoder recommendation method (NARM) to combine sequentiality and user's general preference \cite{li2017neural}. However, STAMP \cite{liu2018stamp} also adopts the concept combining general and current interest, but the difference is that STAMP explicitly models the current interest reflected by the last click to emphasize the importance of the last click, while NARM considers them as equally important. Most recently, geometric deep learning has become popular in a variety of tasks. SR-GNN \cite{wu2019srgnn} transforms sessions into the graph-structured data and applies GNN based on that. The significant improvement in recommendation performance proves the potential of geometric deep learning in SRS, and the motivation behind this work is enlightening to our work. However, all aforementioned deep learning methods only consider item transitions within a single session, which limits the upper bound of performance because of the lack of collaborative information.

Moreover, Collaborative Filtering (CF) idea-based methods are also popular in SRS. Unlike traditional user-based \cite{Jin04} or item-based \cite{Linden03,Billsus98,Pareek13,Sarwar00} CF models in other recommendation tasks, modifications need to be made for them to perform well in SRS. Simply using item neighborhood information \cite{sarwar2001item} cannot extract the integrity and sequentiality of items in the current session, which are extremely important for SRS application scenarios because of the lack of auxiliary data. Thus, SKNN \cite{bonnin2014sknn} is proposed to consider each session as a whole and its improved version KNN-RNN \cite{jannach2017recurrent} integrates GRU4REC to extract the sequentiality. Later, an end-to-end neural model (CSRM \cite{wang2019CSRM}) outperforms KNN-RNN with learnable latent session representations. The major difference between our method and theirs (KNN-RNN and CSRM) is that they only stay at the minimum granularity of session as the collaborative information, while we dig deeper and integrate item transitions among different sessions into SRS.

\subsection{Graph Embedding}
\noindent The most important part of aforementioned deep learning-based methods is embedding because generating more accurate and meaningful session embedding directly decides the performance. Thus, graph embedding becomes a critical component of the method when it comes to graph-structured data. However, traditional kernel-based methods (e.g., Weisfeiler-Lehman kernel \cite{shervashidze11}, Deep Graph Kernels \cite{Yanardag15}) focus more on the unsupervised tasks and have trouble scaling to large graphs, so we mainly introduce neural network-based graph embedding methods here. 

The concept of Graph Neural Networks (GNN) is first purposed by Gori $\textit{et al.}$ \cite{Gori05}, then developed and deepened by Scarselli $\textit{et al.}$ \cite{scarselli09} and Micheli $\textit{et al.}$ \cite{micheli09}. These early methods mainly generate representations of target nodes by using the recurrent neural unit to aggregate information of neighbor nodes. Inspired by the success of Convolutional Neural Network (CNN) in the image classification task, Bruna $\textit{et al.}$ \cite{Bruna14} propose the spectral Graph Convolutional Neural Network (GCN). Then Defferrard $\textit{et al.}$ propose a variant model by introducing fast localized spectral filtering \cite{Defferrard16}, and Kipf $\textit{et al.}$ improve it with a first-order approximation of spectral graph convolutions to motivate the choice of convolutional architecture \cite{Kipf16}. Moreover, Message Passing Neural Network (MPNN) generalizes these GCN-based methods and introduces a two-step framework: message passing and readout \cite{Gilmer17}. Gated Graph Neural Networks (GGNN) \cite{Li15} extends GNN to the sequential output, which is of great significance for sequential recommendations, such as SRS. However, as a large number of studies have shown that the attention mechanism improves the performance of deep learning-based methods in various tasks, it is therefore natural for researchers to import it on graphs \cite{Velickovic17,choi17}. Most recently, Wu $\textit{et al.}$ \cite{wu2019simplifying} discovers that there is lots of unnecessary computation in GCN, and reduces its complexity by developing the simplified GCN (SGCN). After removing nonlinearities and collapsing weight matrices, SGCN can have a better efficiency without hurting the accuracy performance. Thus, we integrate this simplifying approach into our model.

% Velickovic $\textit{et al.}$ \cite{Velickovic17} propose Graph Attention Network (GAT), which uses attention mechanisms to learn node embedding in a graph. By making the weights of nodes trainable, GAT can extract more information from the most critical part of the graph structure without a priori knowledge of structure, which is especially important for the scalability of graph embedding. 

\section{Preliminary}
\label{p}
The goal of SRS can be defined as using users’ current sequential session data to predict users’ next click items. Let $\mathcal{V} = \{v_1, v_2, ... , v_{m}$\} represents a set of unique items in all sessions. 
$s = \{v_1, v_2, ... , v_n\}$ represents an anonymous session which contains items ordered by timestamps. $S = \{s_1, s_2, ... , s_{|S|}\}$ denotes the whole session set. 
For each item in $\mathcal{V}$, we embed it into a unified embedding space. 
Let $\mathbf{v}_i \in \mathbb{R}^{d}$ denotes the latent vector of corresponding item $v_i \in \mathcal{V}$.
% and $ V = \{\mathbf{v}_1, \mathbf{v}_2, ... ,\mathbf{v}_{m}\}$ represents a set of all latent representations. 

Given a session $s$, our model aims to predict the user's possible next click item $v_{n+1}$. We generate probabilities $\widehat{\mathbf{y}}$ for all possible items based on the input target session $s$. Each element's value of vector $\widehat{\mathbf{y}}$ is the recommendation score of the corresponding item. The items with a top-$T$ recommendation score will be recommended as our model's output.

\section{Our Proposed Method: DGTN}
\label{m}
% We here introduce our proposed method DGTN.
% We first talk about how to construct the session graph consisting of both the input current session and its neighbor (similar) sessions.
% Then we elaborate the devised Dual-channel GNN which models both intra- and inter-session transition to learn the item embeddings, the session pooling operation to obtain session embedding, and the prediction layer.

\subsection{Graph Construction}

\begin{figure}[t]
\centering
\includegraphics[width=1\columnwidth]{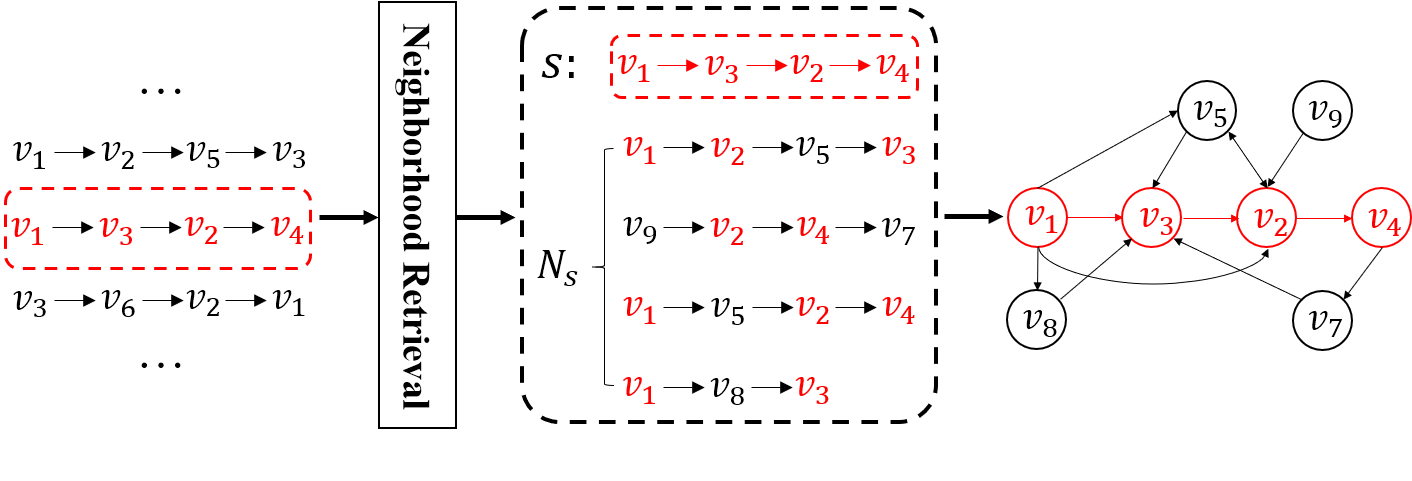} % Reduce the figure size so that it is slightly narrower than the column. Don't use precise values for figure width. This setup will avoid overfull boxes.  
\caption{The construction of the graph. We select the neighbor set $N_s$ from previous sessions based on their similarities to the target session (red) and construct them as a session graph. }
\label{fig:graph}
\vspace{-2mm}
\end{figure}
To model item transitions among different sessions in an explicit manner, we model the sessions in the graph structure (Figure \ref{fig:graph}).
Given a session $s$, we first determine its neighbor set $N_s$ consisting of its $r$ most similar previous sessions. 
To simplify the computation, we use the number of duplicate items between sessions to calculate the similarity. Based on the ranking of the number of duplicate items between each session and the target session, we sample the top-$r$ sessions to constitute the neighborhood set $N_s$.
% To measure the similarities between sessions, we encode each session $s$ as a binary vector $\vec{s} \in \mathbb{R}^{m}$, where if an item appears in the session then the corresponding element in the vector is set to one, otherwise zero. 
% Then we use cosine similarity to measure the similarity between $\vec{s}$ and each session $s_i$ in the previous session set, which can be defined as:
% %First, we compute the cosine similarities between $s$ and every other session $s^' \in \mathcal{S}^{\prime}$. Sessions $s$ and $s^'$ are encoded as binary vectors $\vec{s}, \vec{s^'} \in \mathbb{R}^{|I|}$, where if an item appears then the corresponding element in the vector is set to one, otherwise zero. Then we use cosine similarity to measure the similarity between $\vec{s}$ and $\vec{s_j}$, which can be defined as:
% \begin{equation} \label{eq1}
%     \begin{split}
%     sim(\vec{s},\vec{s_i}) = \frac{\vec{s} \cdot \vec{s_i}}{\sqrt{l(s) \cdot l(s_i)}},
%     \end{split}
% \end{equation}
% where $l(s)$ and $l(s_i)$ represent the length of $s$ and $s_i$, respectively.  
% The top-$r$ similar sessions are selected to constitute the neighborhood set $N_s$. 

Then we model the target session $s$ and its neighbor sessions in $N_s$ as a session graph $\mathcal{G}_{s}$, where each node represents either an item that appears in the session $s$ or any neighbor session in $N_s$.
Each edge $(v_i, v_j)$ denotes a user clicking on item $v_{j}$ after $v_{i}$ in the session $s$ or any neighbor session in $N_s$. 
We denote $\mathcal{V}_{s}$ as the set containing items in session $s$. And $\mathcal{V}_{N_s}$ denotes the set of items belonging to $N_s$.

\begin{figure}[t]
\centering
\includegraphics[width=\columnwidth]{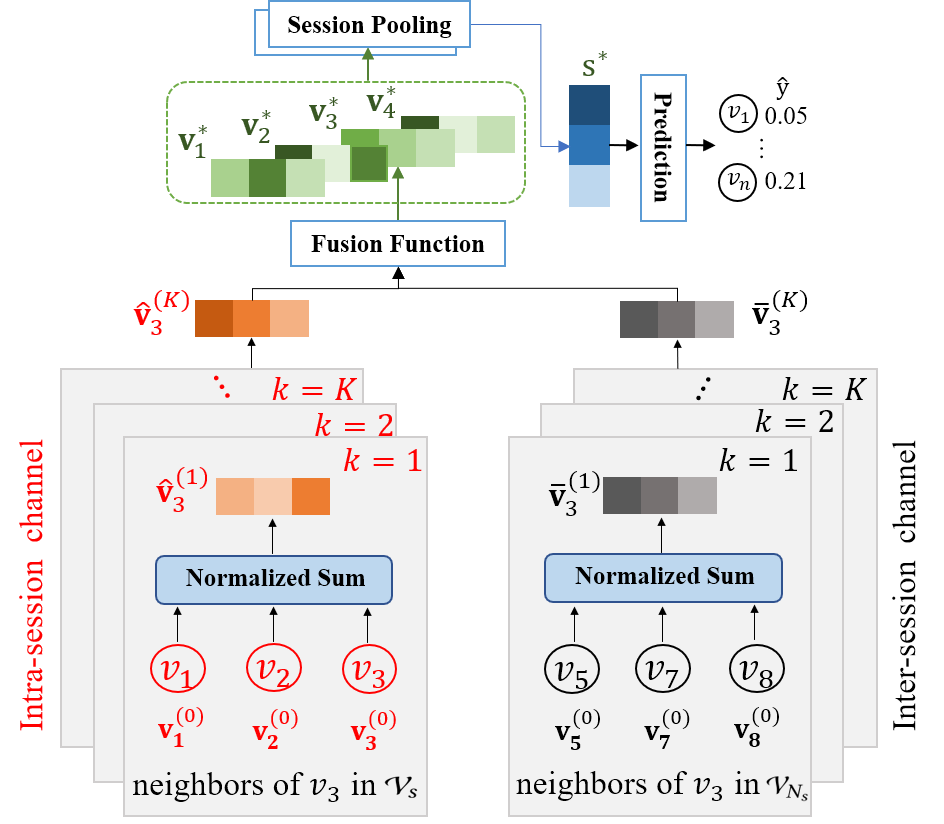} % Reduce the figure size so that it is slightly narrower than the column. Don't use precise values for figure width. This setup will avoid overfull boxes. 
\caption{An illustration of the model architecture. Taking item $v_3$ for example, we propagate the embeddings of neighbor items in the target session $s$ and the neighbor session set $N_s$ towards the next layer in the intra- and inter-session channel, respectively. Then the embeddings of the final layer in the two channels are fed into the fusion function to obtain the final item embedding $\mathbf{v}^{*}_3$. Based on the learned item embeddings, we generate the session embedding via session pooling. Finally, we apply a prediction layer to generate the recommendation probability $\hat{\mathbf{y}}$. 
}
\label{fig:embedding}
\vspace{-2mm}
\end{figure}

\subsection{Item Embedding Learning}
% The constructed graph contains item nodes from both target session $s$ and neighbor sessions in $N_s$, which reflect the users' current behaviors and the global collaborative information, respectively. Therefore, they ought to have different effects on the recommendation. We thus propagate the item embedding in two channels to model the inherent differences between these two types of nodes.
% In the intra-session channel, we aggregate messages from only the neighbor nodes in $\mathcal{V}_s$, and the updated set of item embeddings is denoted as $\hat{\mathbf{V}}_s = \{\hat{\mathbf{v}}_1, \hat{\mathbf{v}}_2, ..., \hat{\mathbf{v}}_n \}$.
% By contrast, in the inter-session channel, we aggregate messages from the neighbor nodes in $\mathcal{V}_{N_s}$ and obtain the set of item embeddings $\overline{\mathbf{V}}_s = \{\overline{\mathbf{v}}_1, \overline{\mathbf{v}}_2, ..., \overline{\mathbf{v}}_n \}$.
% Following the message-passing structure of GNN~\cite{kipf2016semi}, each item node $v_i$ in the graph aggregates messages passed from its neighbor item nodes in the graph. Inspired by SGCN \cite{wu2019simplifying}, we remove the nonlinearities and collapse the weight matrix into one weight matrix. Only the normalized sum of neighbor embeddings are propagated towards next layers:
To encode the item transition signals into item embeddings, we follow the message-passing structure of GNN \cite{kipf2016semi}. Each item node $v_i$ in the graph aggregates messages passed from its neighbor item nodes in the graph.
However, the constructed session graph contains item nodes from target session $s$ and neighbor sessions $N_s$, which reflect the users' current behaviors and the global collaborative information, respectively.
% However, neighbor item nodes from target session $s$ and neighbor sessions $N_s$ are, which reflect the users' current behaviors and the global collaborative information, respectively.
They ought to have different effects on the recommendation.
Therefore, we propagate the item embeddings in two channels for the two types of neighbor item nodes.
In the intra-session channel, we aggregate messages from only the neighbor nodes in $\mathcal{V}_s$.
In the inter-session channel, we aggregate messages from the neighbor nodes in $\mathcal{V}_{N_s}$.
% and the updated set of item embeddings is denoted as $\hat{\mathbf{V}}_s = \{\hat{\mathbf{v}}_1, \hat{\mathbf{v}}_2, ..., \hat{\mathbf{v}}_n \}$.
%  and obtain the set of item embeddings $\overline{\mathbf{V}}_s = \{\overline{\mathbf{v}}_1, \overline{\mathbf{v}}_2, ..., \overline{\mathbf{v}}_n \}$.
Inspired by SGCN \cite{wu2019simplifying}, we remove the nonlinearities and collapse the weight matrix into one weight matrix. Only the normalized sum of neighbor embeddings are propagated towards next layers:
\begin{equation}\begin{aligned}
\hat{\mathbf{v}}_{i}^{(k)} &= \frac{1}{d_{i}+1} \mathbf{v}_{i}^{(k-1)}+\sum_{v_{j} \in \mathcal{V}_{s}} \frac{a_{i j}}{\sqrt{\left(d_{i}+1\right)\left(d_{j}+1\right)}} \mathbf{v}_{j}^{(k-1)}, \\
\overline{\mathbf{v}}_{i}^{(k)} &= \frac{1}{d_{i}+1} \mathbf{v}_{i}^{(k-1)}+\sum_{v_{j} \in \mathcal{V}_{N_s}} \frac{a_{i j}}{\sqrt{\left(d_{i}+1\right)\left(d_{j}+1\right)}} \mathbf{v}_{j}^{(k-1)}, 
\end{aligned}\end{equation}
where $d_{i}$ is the degree of node $v_{i}$ in the adjacency matrix. $a_{ij} = 1$ denotes that there is an edge between node ${v_i}$ and $v_{j}$, and a missing edge is represented through $a_{ij} = 0$.
The updated set of item embeddings from the intra-session channel is denoted as $\hat{\mathbf{V}}_s = \{\hat{\mathbf{v}}_1, \hat{\mathbf{v}}_2, ..., \hat{\mathbf{v}}_n \}$, where $\hat{\mathbf{v}}_i$ = $\hat{\mathbf{v}}_i^{(K)}$ and $K$ is the number of layers.
By contrast, the set from the inter-session channel is denoted as $\overline{\mathbf{V}}_s = \{\overline{\mathbf{v}}_1, \overline{\mathbf{v}}_2, ..., \overline{\mathbf{v}}_n \}$.
Likewise, $\overline{\mathbf{v}}_i$ = $\overline{\mathbf{v}}_i^{(K)}$. 
% The constructed graph contains item nodes from both target session $s$ and neighbor sessions in $N_s$, which reflect the users' current behaviors and the global collaborative information, respectively. Therefore, they ought to have different effects on the recommendation. We thus propagate the information in different channels to model the inherent differences between different types of nodes. We update two set of item embeddings $\hat{\mathbf{V}}_s = \{\hat{\mathbf{v}}_1, \hat{\mathbf{v}}_2, ..., \hat{\mathbf{v}}_n \}$, $\overline{\mathbf{V}}_s = \{\overline{\mathbf{v}}_1, \overline{\mathbf{v}}_2, ..., \overline{\mathbf{v}}_n \}$ by aggregating messages from only the nodes in $\mathcal{V}_s$ and all nodes including those in $N_s$, respectively. And then we concatenate them into a representation matrix $\mathcal{M}_{i} \in \mathbb{R}^{2 \times d}$.
% \begin{equation} \mathcal{M}_i = \hat{\mathbf{v}}_i \| \overline{\mathbf{v}}_i\end{equation}

Then we need to fuse the information of these two channels together and extract effective features (Figure \ref{fig:embedding}). Here we explore five different fusion functions: (a) Mean Pooling (Mean) fuses the intra- and inter-session channel information by taking the mean value of every dimension of two embeddings.  (b) Max Pooling (Max) takes the maximum value of every dimension of two embeddings. (c) Concatenation (Concat) is the concatenation of two item embeddings. (d) Fusion Gating (FG) is a linear interpolation between the two item embeddings $\hat{\mathbf{v}}_{i}$ and $\overline{\mathbf{v}}_{i}$, inspired from the session fusion gating in CSRM:
\begin{equation}\begin{aligned}
\mathbf{f} &=\sigma\left(\mathbf{W}_{f}^{1} \hat{\mathbf{v}}_{i}+\mathbf{W}_{f}^{2} \overline{\mathbf{v}}_{i}+\mathbf{b}_{f}\right), \\
\mathbf{v}_{i}^{*} &=\mathbf{f}\hat{\mathbf{v}}_{i}+(1-\mathbf{f}) \overline{\mathbf{v}}_{i},
\end{aligned}\end{equation}
where $\mathbf{W}_{f}^1$, $\mathbf{W}_{f}^2$ denote weight matrices in fusion gate and $\mathbf{b}_{f}$ denotes the bias vector. $\mathbf{v}^*_i$ is the final embedding for item $v_i$. (e) Convolution Neural Networks (CNN) Pooling is the strategy employed in our method, which will be elaborated later.
Among multiple fusion functions, we employ CNN pooling according to experimental results. Suppose we have $f$ vertical convolution filters, $\boldsymbol{F}^{k} \in \mathbb{R}^{2 \times 1}, 1 \leq k \leq f$. Each of them interacts with the columns of $\mathcal{M}_{i}$, the concatenation of two item embeddings, by sliding from left to right. Then the results of $f$ convolution filters are concatenated and transformed to the final item embedding $\mathbf{v}_{i}^{*}$:
\begin{equation}\begin{aligned}
\mathcal{M}_i &= \hat{\mathbf{v}}_i^{(K)} \| \overline{\mathbf{v}}_i^{(K)}, \\
\mathbf{v}_{i}^{*} &= \mathbf{W}_{i}\left(c^{1}\left\|c^{2}\right\| \ldots \| c^{f}\right),
\end{aligned}\end{equation}
where $\mathcal{M}_{i} \in \mathbb{R}^{2 \times d}$ is the concatenation of two item embeddings, $c^{k} \in \mathbb{R}^{1 \times d}, 1 \leq k \leq f$ is the convolution result of filter $\boldsymbol{F}^{k}$, $\mathbf{W}_{i} \in \mathbb{R}^{1 \times f}$ projects $o_{i} \in \mathbb{R}^{f \times d}$ into embedding space $\mathbb{R}^{d}$.

\subsection{Session Pooling}
Inspired by STAMP, we combine both users' long-term preference and short-term interests of the session to generate the final session embedding.
For the session $s = \{v_1, v_2, ... , v_n\}$, we use the last click item's embedding to represent user's short-term interests, i.e., $\mathbf{p}_{s} = \mathbf{v}^*_n$. Then we obtain the long-term preference $\mathbf{p}_{l}$ by adopting a soft-attention mechanism to draw dependencies between the short-term interest $\mathbf{p}_{s}$ and each item in the session. Specifically, we derive $\mathbf{p}_{l}$ by the following calculation:
\begin{equation} \label{eq5}
    \begin{split}
    \begin{aligned}
    \alpha_{i} &= \operatorname{softmax}\left(\mathbf{q}^{T} \left(\mathbf{W}^{1}_a \mathbf{p}_{s}+\mathbf{W}^{2}_a \mathbf{v}^{*}_{i}+\mathbf{W}^{3}_a \mathbf{v}^{*}_{avg}+\mathbf{b}_a \right)\right), \\
    \mathbf{p}_{l} &= \sum\nolimits_{i=1}^{n} \alpha_{i} \mathbf{v}^{*}_{i},
    \end{aligned}
    \end{split}
\end{equation}
% \begin{equation} \label{eq5}
%     \begin{split}
%     \begin{array}{l}{\alpha_{i} = \mathbf{q}^{T} \sigma\left(\mathbf{W}^{1}_a \mathbf{v}^{*}_{n}+\mathbf{W}^{2}_a \mathbf{v}^{*}_{i}+\mathbf{b}_a \right)}, \\
%     {\mathbf{s}_{\mathrm{l}} = \sum_{i=1}^{n} \alpha_{i} \mathbf{v}^{*}_{i}},\end{array}
%     \end{split}
% \end{equation}
where $\mathbf{q}^{T} \in \mathbb{R}^{d}$ is a projection vector, $\mathbf{W}^{1}_a, \mathbf{W}^{2}_a \in \mathbb{R}^{d \times d}$ are learnable weighted parameters, $\mathbf{v}^{*}_{avg} = \sum_i \left(\mathbf{v}^{*}_{i}\right) / n$ is the average item embedding, and $\mathbf{b}_a$ is a bias vector.

Finally, we combine $\mathbf{p}_{l}$ and $\mathbf{p}_{s}$ to generate the final session embedding $\mathbf{s}^{*}$:
\begin{equation} \label{eq6}
    \begin{split}
    \mathbf{s}^{*}=\mathbf{W}_c\left[\mathbf{p}_{l} || \mathbf{p}_{s}\right],
    \end{split}
\end{equation}
where $\mathbf{W}_c \in \mathbb{R}^{d \times 2 d}$ transfers the concatenation vector from latent space $\mathbb{R}^{2d}$ to $\mathbb{R}^{d}$.

% \begin{table*}
% \centering
% \scriptsize
% \caption{Statistics of datasets used in our experiments.}\label{tab1}
% \resizebox{0.9\textwidth}{!}{
% \begin{tabular}{l ccccc}
%     \toprule
%     \textbf{Datasets} & \# of clicks &  \# of training sessions &  \# of testing sessions  &  \# of items & average length \\
%     \midrule

%     \textbf{Yoochoose 1/64}  
%              &   565,552   &   375,043  &   55,405   &   17,319   &   6.07           \\
%     \textbf{Yoochoose 1/4}     
%              &   7,980,529   &   5,969,416  &   55,872   &   30,638   & 5.71            \\
%     \textbf{Diginetica}    
%              &   982,961   &   719,470  &   68,977   &   43,097   &   5.12           \\
    
%     \bottomrule
% \end{tabular}}
% \end{table*}

\subsection{Prediction Module and Objective Function}
After obtaining the final session representation $\mathbf{s}^{*}$, we use it to multiply each candidate item vector $\mathbf{v}_{\mathrm{i}}$ to generate recommendation score $\hat{\mathbf{z}_{i}}$ for corresponding item:
\begin{equation} \label{eq14}
    \begin{split}
    \hat{\mathbf{z}}_{i}=\mathbf{v}_{i}^\top \mathbf{s}^{*}.
    \end{split}
\end{equation}
Then we apply a softmax function to generate the output vector $\hat{\mathbf{y}}$ of the model:
\begin{equation} \label{eq15}
    \begin{split}
    \hat{\mathbf{y}}=\operatorname{softmax}(\hat{\mathbf{z}}),
    \end{split}
\end{equation}
where $\hat{\mathbf{z}} \in \mathbb{R}^{m}$ represents the recommendation scores over all candidate items, and $\hat{\mathbf{y}} \in \mathbb{R}^{m}$ denotes the probabilities of items becoming the next-click item in session $s$.

In the training process, we apply cross-entropy as the loss function:
\begin{equation} \label{eq16}
    \begin{split}
    \mathcal{L}(\hat{\mathbf{y}})=-\sum_{i=1}^{m} \mathbf{y}_{i} \log \left(\hat{\mathbf{y}}_{i}\right)+\left(1-\mathbf{y}_{i}\right) \log \left(1-\hat{\mathbf{y}}_{i}\right),
    \end{split}
\end{equation}
where $\mathbf{y}$ denotes the one-hot encoding vector of the ground truth item.
Finally, our model is trained by Back-Propagation Through Time (BPTT) algorithm.

\section{Experiments}
\label{e}
In this section, we aim to answer the following research questions:
\begin{itemize}
    \item {\textbf{RQ1.} How does DGTN perform as compared with state-of-the-art SRS methods?}
    \item {\textbf{RQ2.} How does the number of neighbor sessions affect the model performance?}
    \item {\textbf{RQ3.} How do different fusion functions affect the model performance?}
    \item {\textbf{RQ4.} What is the training efficiency of DGTN?}
\end{itemize}

\begin{table*}
\centering
\scriptsize
\caption{Statistics of datasets used in our experiments.}\label{tab1}
\resizebox{0.9\textwidth}{!}{
\begin{tabular}{l ccccc}
    \toprule
    \textbf{Datasets} & \# of clicks &  \# of training sessions &  \# of testing sessions  &  \# of items & average length \\
    \midrule

    \textbf{YOOCHOOSE 1/64}  
             &   565,552   &   375,043  &   55,405   &   17,319   &   6.07           \\ [0.5ex]

    \textbf{Diginetica}    
             &   982,961   &   719,470  &   68,977   &   43,097   &   5.12           \\
    
    \bottomrule
\end{tabular}}
\end{table*}

We conduct experiments on two real-world datasets: Yoochoose\footnote{http://2015.recsyschallenge.com/challenge.html} and Diginetica\footnote{http://cikm2016.cs.iupui.edu/cikm-cup}. 
We apply the same preprocessing as \cite{liu2018stamp, wu2019srgnn}.
We filter out sessions of length one and items that appear less than five times for both datasets as same as previous studies \cite{liu2018stamp, wu2019srgnn}. Furthermore, we use the last one day in YOOCHOOSE and last seven days in Diginetica to generate the test data. The data stastic is shown in Table \ref{tab1}. Because collaborative filtering-based methods cannot recommend an item which has not appeared before \cite{hidasi2015session}, we filter out items from test set which do not appear in the training set. Following previous studies \cite{liu2018stamp, wu2019srgnn}, We only use the most recent 1/64 of the training sequence of Yoochoose. We use Recall@20 and MRR@20 as evaluation metrics.

\textbf{Baseline. }
We compare DGTN with frequency based method (POP), two RNN-based methods (GRU4REC \cite{hidasi2015session} and NARM \cite{li2017neural}), attention-based method (STAMP \cite{liu2018stamp}), two traditional Matrix Factorization or Markov Chain approaches (FPMC \cite{rendle2010factorizing} and BPR-MF \cite{rendle2009bpr}), and GNN-based method (SR-GNN \cite{wu2019srgnn}). We also compare the model with some neighborhood-based methods. Although they consider the collaborative information among multiple sessions, they either ignore the sequentiality of items (Item-KNN \cite{sarwar2001item} and SKNN \cite{bonnin2014sknn}) or only consider coarse-grained session-level collaborative information (KNN-RNN \cite{jannach2017recurrent} and CSRM \footnote{Following previous studies \cite{liu2018stamp, wu2019srgnn}, we set the hidden size of all baseline to 100, which is different from the original setting of CSRM.} \cite{wang2019CSRM}), failing to model the complex item transitions between different sessions.

\begin{table}
\centering
\scriptsize
\caption{Performance comparison of different methods.}\label{tab2}
\vspace{-3mm}
\resizebox{0.45\textwidth}{!}{
\begin{tabular}{ccccc}
% \begin{tabular}{cllllllllllll}
        \toprule
    \multirow{3}{*}{\textbf{Methods}} & \multicolumn{2}{c}{\textbf{Diginetica}} & \multicolumn{2}{c}{\textbf{Yoochoose 1/64}}\\
    \cmidrule(r){2-3} \cmidrule(r){4-5} 
    
     & \textbf{MRR@20} & \textbf{Recall@20} & \textbf{MRR@20} & \textbf{Recall@20} \\
    
     \midrule
                 %Mrr             % Recall
              % 10     20         10    20   
    \textbf{POP}      
             & 0.20       & 0.89 
             & 1.65       & 6.71   \\
    \textbf{Item-KNN}    
             & 11.57      & 35.75  
             & 21.82      & 51.60  \\
    \textbf{SKNN} 
             & 16.77      & 45.45  
             & 24.03      & 57.27  \\
    \textbf{FPMC}     
             & 6.95      & 26.53 
             & 15.01      & 45.62  \\        
    \textbf{BPR-MF}   
             & 1.98      & 5.24  
             & 12.08      & 31.31  \\
    \textbf{GRU4REC}  
             & 8.33      & 29.45  
             & 22.89      & 60.64  \\
    \textbf{NARM} 
             & 16.17      & 49.70  
             & 28.63      & 68.32  \\
    \textbf{STAMP}    
             & 14.32      & 45.64 
             & 29.67      & 68.74  \\
    \textbf{KNN-RNN}  
             & 9.65       & 31.89 
             & 24.05      & 62.36  \\
    \textbf{CSRM}
             & 17.16      & 52.00
             & 30.48      & 70.79  \\
    \textbf{SR-GNN}   
             & 17.59      & 50.73
             & 30.94      & 70.57  \\
    \midrule
    % $\textbf{DGTN}_\text{intra}$    
    %          & 30.27     & 60.52  
    %          & 31.19      & 61.34
    %          & 27.13      & 52.97  \\
    \textbf{DGTN}    
             & \textbf{18.07}      & \textbf{53.05}
             & \textbf{31.35}      & \textbf{71.18}  \\
    \textbf{Improv.(\%)} 
             & 2.72      & 2.02
             & 1.33      & 0.56  \\
    \bottomrule
\end{tabular}}
\vspace{-5mm}
\end{table}

\subsection{Comparison with baseline methods (RQ1)}
The experimental results of all methods are illustrated in Table \ref{tab2}, and the following observations stand out:
\begin{itemize}
    \item {Two KNN-based methods, Item-KNN and SKNN, consistently achieve better performance than other conventional methods. Both RNN-based methods (GRU4REC and NARM) underperform their neighborhood-enhanced versions (KNN-RNN and CSRM). This illustrates the effectiveness of adopting collaborative information from other sessions.}
    \item {All of the neural network-based methods distinctly outperform other conventional recommendation methods, demonstrating the superiority of adopting deep learning technology to make recommendations. The key reason for this may be RNN’s ability to process sequentiality and thus model the item transitions within the target session.}
    \item {On the whole, graph-based methods (SR-GNN, and DGTN) outperform RNN-based methods (GRU4REC, NARM, KNN-RNN, and CSRM). This indicates the significance of explicitly modeling complex contextual item transitions owing to the strong power of GNN in modeling graph-structured data. On the contrary, RNN can only deal with unidirectional transitions between consecutive items.}
    % \item { $\mathrm{DGTN}_\mathrm{intra}$, which only model intra-session item interactions, performs worse than the full DGTN model. It proves the significance of our introduced inter-session item interaction modeling (RQ2).}
    \item {DGTN consistently yields the best performance on all datasets (RQ1), which verifies the superiority of the proposed method. By propagating embedding on the constructed graph, DGTN is capable of exploring complex item transitions among different sessions. And the improvement over SR-GNN indicates the significance of that.} 

    % DGTN shows a significant improvement compared with the neighborhood-based methods (CSRM and KNN-RNN), which verifies the superiority of our designed DGNN in modeling inter-session item interactions.}
\end{itemize}

\begin{figure}[t]
\centering
\subfigure[Number of neighbor sessions]{
\begin{minipage}[b]{0.23\textwidth}
%\centering
\label{fig:neighbor} %% label for first subfigure
\includegraphics[width=1\textwidth]{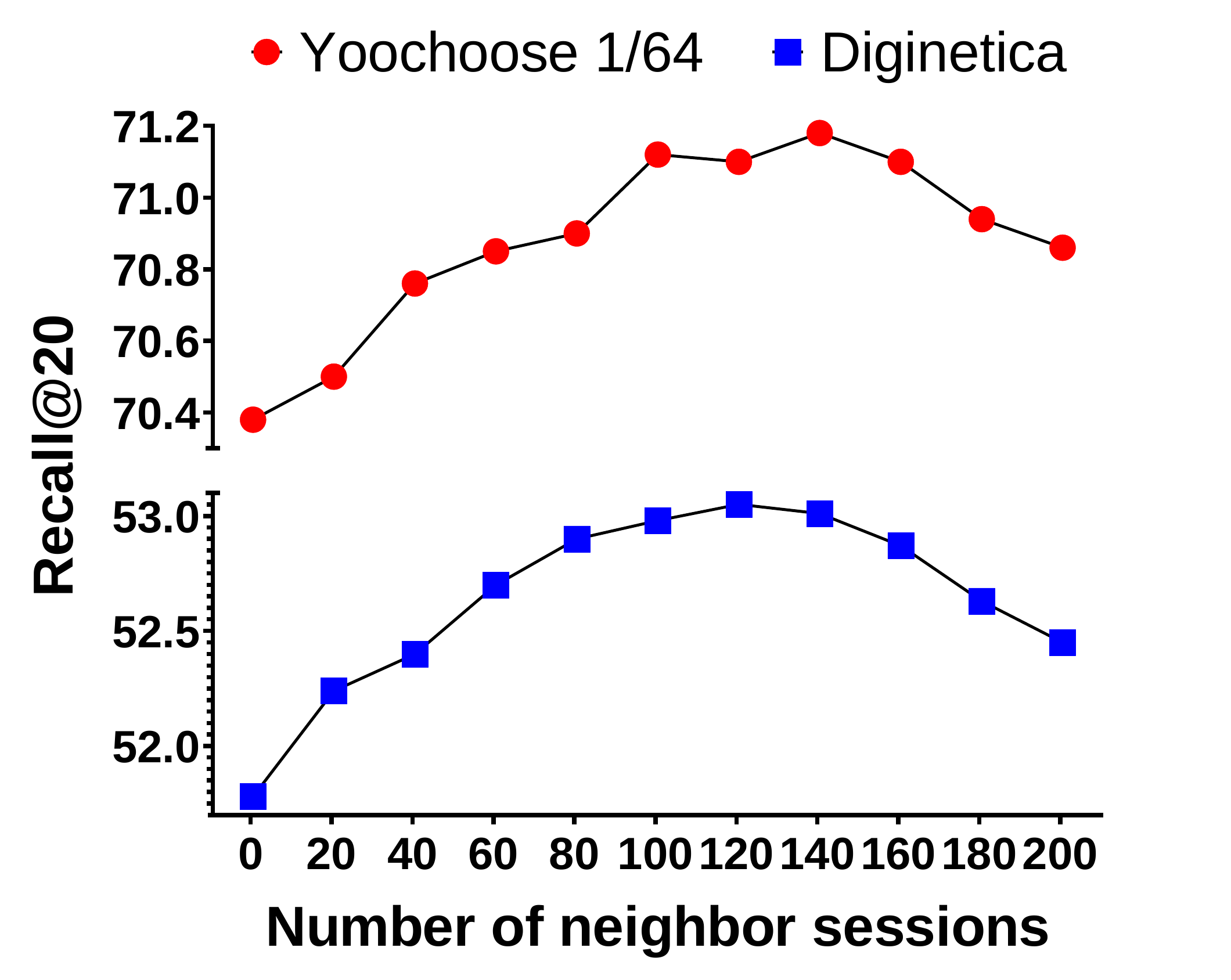}
\end{minipage}%
}
\subfigure[Different fusion functions]{
\begin{minipage}[b]{0.23\textwidth}
%\centering
\label{fig:layer} %% label for first subfigure
\includegraphics[width=1\textwidth]{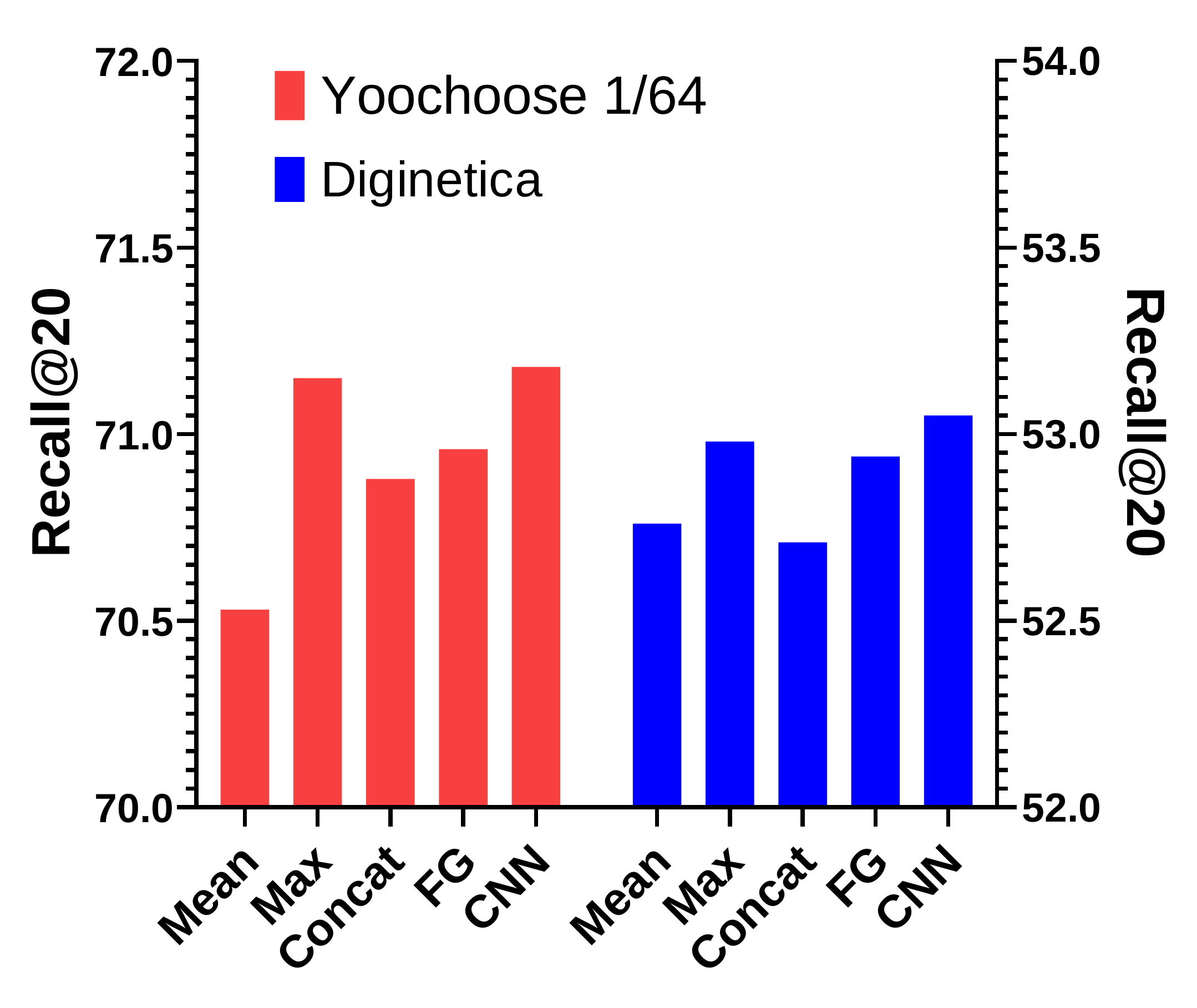}
\end{minipage}%
}%
\vspace{-4mm}
\caption{Performance w.r.t. number of neighbor sessions (a), and fusion functions (b). }
\label{fig:performance}
\vspace{-4mm}
\end{figure}

\subsection{Effect of the number of neighbors (RQ2)}
Focusing on the current session $s$ allows the model to dig deeper into the current user preference, but loses valuable collaborative information with other sessions; Introducing a large set of neighbor sessions $N_s$ widens the range of information but leads to more noises. Thus, we vary the number of neighbor sessions $r$ from zero\footnote{$r=0$ means that the model only consider transitions within the target session.} to 200, to study the trade-off between them (RQ2). In Figure~\ref{fig:neighbor}, the performance of DGTN improves with the increase of $r$ at first since more collaborative information is introduced. However, it starts to drop after $r = 140$ on Yoochoose 1/64, and $r = 120$ on Diginetica, because of the increasing noise introduced by less similar sessions (the similarity decreases as $r$ increases). Differences in the optimal numbers might be caused by the statistical differences between two datasets. And it is clear that neighbor sessions significantly increase the model performance, which verifies the rationality and effectiveness of integrating dual-channel item transitions.
% \begin{table}
% \centering
% \scriptsize
% \caption{Performance comparison of different fusion functions.}\label{tab2}
% \vspace{-3mm}
% \resizebox{0.45\textwidth}{!}{
% \begin{tabular}{ccccccc}
% % \begin{tabular}{cllllllllllll}
%         \toprule
%     \multirow{4}{*}{\textbf{Methods}} & \multicolumn{2}{c}{\textbf{Yoochoose 1/64}} & \multicolumn{2}{c}{\textbf{Yoochoose 1/4}} & \multicolumn{2}{c}{\textbf{Diginetica}}\\
%     \cmidrule(r){2-3} \cmidrule(r){4-5} \cmidrule(r){6-7} 
    
%      & \textbf{MRR} & \textbf{Recall} & \textbf{MRR} & \textbf{Recall} & \textbf{MRR} & \textbf{Recall} \\
    
%      \midrule
%                  %Mrr             % Recall
%               % 10     20         10    20   

%     \textbf{Average Pooling}    
%              & -     & -  
%              & -      & -
%              & -      & -  \\
%     \textbf{Max Pooling}    
%              & -     & -  
%              & -      & -
%              & -      & -  \\
%     \textbf{Concatenation}    
%              & -     & -  
%              & -      & -
%              & -      & -  \\
%     \textbf{Fusion Gating}    
%              & -     & -  
%              & -      & -
%              & -      & -  \\             
%     \textbf{DGTN}    
%              & \textbf{30.61}      & \textbf{60.88}  
%              & \textbf{31.50}      & \textbf{61.77} 
%              & \textbf{27.46}      & \textbf{53.26}  \\
    
%     \bottomrule
% \end{tabular}}
% \vspace{-5mm}
% \end{table}
\subsection{Effect of fusion functions (RQ3)}
To analyze how different fusion functions affect the results (RQ3), we test the model with five different fusion functions. From Figure~\ref{fig:performance}, observations of the results can be listed as follows:
\begin{itemize}
    \item {CNN pooling outperforms others on both datasets in terms of Recall@20. And the same phenomenon occurs in the MRR@20 evaluation, which is omitted for the space limit. This indicates that CNN has a stronger ability to fuse intra- and inter-session channel features, which might be attributed to its representational capacity.}
    \item {Mean Pooling and Concatenation show bad performance in both datasets. The reason may be that a simple overall smoothness does not well deal with the differences between two different channels for the diversity of different user behaviors.}
    \item {Max Pooling is also a good choice because of its relatively high performance and computational efficiency, which illustrates the effectiveness of capturing the most important feature.}
\end{itemize}

\subsection{Efficiency (RQ4)}

\begin{table}[htb]
\caption{Efficiency (training time per epoch and GPU memory usage) comparison between SR-GNN and DGTN.}
\label{tab:eff}
\resizebox{0.45\textwidth}{!}{
\renewcommand{\arraystretch}{1.3}
\begin{tabular}{ccccc}
\toprule
\multirow{2}{*}{Methods} & \multicolumn{2}{c}{Diginetica} & \multicolumn{2}{c}{Yoochoose 1/64} \\ \cline{2-5} 
       & Time (s) & Memory (MB) & Time (s) & Memory (MB) \\ \midrule
SR-GNN & 601      & 973         & 522      & 961         \\
DGTN   & 895      & 1031         & 679      & 995         \\ \bottomrule
\end{tabular}}
\end{table}

We evaluate the training efficiency of DGTN in this section. As DGTN trains on a larger session graph than previous graph-based methods, we are curious about its efficiency compared with SR-GNN. To make a fair comparison, we set the batch size as 128 and the hidden size as 100 for both methods, following the common implementation \cite{liu2018stamp, wu2019srgnn}. All experiments are conducted on a single GeForce RTX 2080ti GPU and the same computation environment. Both methods are training with 15 epochs and we report the average training time per epoch. And the GPU memory usages are recorded when the training stabilizes. The results of the training time per epoch and the GPU memory usage are shown in Table \ref{tab:eff}.

From Table \ref{tab:eff}, we can observe that DGTN performs worse than SR-GNN, which is reasonable as DGTN uses a bigger session graph. However, the difference between GPU memory usage is minor and the training time of DGTN is also acceptable considering the performance improvement.

\section{Conclusion}
\label{c}
We proposed a novel method, DGTN, to explicitly model item transitions among different sessions. We constructed the complex transitions between items among different sessions into a single graph. Then the transition signals are injected into the embedding in the channel-aware learning process. Extensive experiments on two real-world datasets demonstrate the effectiveness of DGTN.

\section*{Acknowledgements}
This work is supported by National Natural Science Foundation of China (61772528, U19B2038) and National Key Research and Development Program (2016YFB1001000, 2018YFB1402600).

\bibliographystyle{ieeetr}
\bibliography{sample-base}

\end{document}